\documentclass[conference]{IEEEtran}

\usepackage[nolist]{acronym}
\usepackage{cite}
\usepackage{amsmath,amssymb,amsfonts}
\usepackage{algorithmic}
\usepackage{graphicx}
\usepackage{textcomp}
\usepackage{xcolor}

\usepackage{comment}
\usepackage{hyperref}
\usepackage{listings}
\usepackage{booktabs}
\usepackage{subcaption}
\usepackage{caption}
\usepackage{balance}
\usepackage{wrapfig}
\usepackage{paralist}
\usepackage{verbatim}

\newif\iffinal

\finaltrue

\iffinal
  \newcommand\ian[1]{}
  \newcommand\kyle[1]{}
  \newcommand\ryan[1]{}
  \newcommand\jim[1]{}
  \newcommand\valerie[1]{}
  \newcommand\justin[1]{}
\else
  \newcommand\ian[1]{{\textcolor{red}{Ian: #1}}}
  \newcommand\kyle[1]{{\textcolor{blue}{Kyle: #1}}}
  \newcommand\ryan[1]{{\textcolor{green}{Ryan: #1}}}
  \newcommand\jim[1]{{\textcolor{purple}{Jim: #1}}}
  \newcommand\valerie[1]{{\textcolor{orange}{Valerie: #1}}}
  \newcommand\justin[1]{{\textcolor{brown}{Justin: #1}}}
\fi



\usepackage{listings}

\usepackage{enumitem}
\setlist[itemize]{leftmargin=*}

\definecolor{delim}{RGB}{20,105,176}
\definecolor{numb}{RGB}{106, 109, 32}
\definecolor{string}{rgb}{0.64,0.08,0.08}

\lstdefinelanguage{json}{
    escapechar=\%,
    breaklines=true,
    postbreak=\raisebox{0ex}[0ex][0ex]{\ensuremath{\color{gray}\hookrightarrow\space}},
    basicstyle=\ttfamily\small,
    upquote=true,
    morestring=[b]",
    stringstyle=\color{string},
    keywords={from,import},
    literate=
     *{0}{{{\color{numb}0}}}{1}
      {1}{{{\color{numb}1}}}{1}
      {2}{{{\color{numb}2}}}{1}
      {3}{{{\color{numb}3}}}{1}
      {4}{{{\color{numb}4}}}{1}
      {5}{{{\color{numb}5}}}{1}
      {6}{{{\color{numb}6}}}{1}
      {7}{{{\color{numb}7}}}{1}
      {8}{{{\color{numb}8}}}{1}
      {9}{{{\color{numb}9}}}{1}
      {\{}{{{\color{delim}{\{}}}}{1}
      {\}}{{{\color{delim}{\}}}}}{1}
      {[}{{{\color{delim}{[}}}}{1}
      {]}{{{\color{delim}{]}}}}{1},
}

\lstdefinelanguage{numberedflow}{
    frame=tb,
    escapechar=\%,
    breaklines=true,
    postbreak=\raisebox{0ex}[0ex][0ex]{\ensuremath{\space}},
    basicstyle=\ttfamily\small,
    upquote=true,
    morestring=[b]",
    stringstyle=\color{string},
    literate=
     *{0}{{{\color{numb}0}}}{1}
      {1}{{{\color{numb}1}}}{1}
      {2}{{{\color{numb}2}}}{1}
      {3}{{{\color{numb}3}}}{1}
      {4}{{{\color{numb}4}}}{1}
      {5}{{{\color{numb}5}}}{1}
      {6}{{{\color{numb}6}}}{1}
      {7}{{{\color{numb}7}}}{1}
      {8}{{{\color{numb}8}}}{1}
      {9}{{{\color{numb}9}}}{1}
      {\{}{{{\color{delim}{\{}}}}{1}
      {\}}{{{\color{delim}{\}}}}}{1}
      {[}{{{\color{delim}{[}}}}{1}
      {]}{{{\color{delim}{]}}}}{1},
}

\lstdefinelanguage{bash}{
    frame=tb,
    escapechar=\%,
    breaklines=true,
    postbreak=\raisebox{0ex}[0ex][0ex]{\ensuremath{\space}},
    basicstyle=\ttfamily\small,
    upquote=true,
    morestring=[b]",
    stringstyle=\color{string},
    literate=
     *{0}{{{\color{numb}0}}}{1}
      {1}{{{\color{numb}1}}}{1}
      {2}{{{\color{numb}2}}}{1}
      {3}{{{\color{numb}3}}}{1}
      {4}{{{\color{numb}4}}}{1}
      {5}{{{\color{numb}5}}}{1}
      {6}{{{\color{numb}6}}}{1}
      {7}{{{\color{numb}7}}}{1}
      {8}{{{\color{numb}8}}}{1}
      {9}{{{\color{numb}9}}}{1}
      {\{}{{{\color{delim}{\{}}}}{1}
      {\}}{{{\color{delim}{\}}}}}{1}
      {[}{{{\color{delim}{[}}}}{1}
      {]}{{{\color{delim}{]}}}}{1},
}

\begin{acronym}
  \acro{BPExxx}{Braid Policy Engine}
  \acro{ML}{machine learning}
  \acro{FAIR}{findable, accessible, interoperable, and reusable}
  \acro{APS}{Advanced Photon Source}
  \acro{ALCF}{Argonne Leadership Computing Facility}
  \acro{SSRL}{Stanford Synchrotron Radiation Lightsource}
  \acro{SSX}{Serial Synchrotron Crystallography}
  \acro{XPCS}{X-Ray Photon Correlation Spectroscopy}
  \acro{HEDM}{High Energy Diffraction Microscopy}
  \acro{AP}{Action Provider}
  \acro{DAG}{Directed Acyclic Graph}
  \acro{ORM}{Object Relational Model}
  \acro{AWS}{Amazon Web Services}
  \acro{ECS}{Elastic Container Service}
  \acro{vCPU}{Virtual CPU}
  \acro{ACU}{Aurora Capacity Unit}
  \acro{CLI}{Command Line Interface}
  \acro{SDK}{Software Development Kit}
  \acro{CRUD}{Create, Read, Update, Delete}
  \acro{SEM}{Scanning Electron Microscope}
\end{acronym}

\def\BibTeX{{\rm B\kern-.05em{\sc i\kern-.025em b}\kern-.08em
    T\kern-.1667em\lower.7ex\hbox{E}\kern-.125emX}}

\begin{document}

\title{Steering a Fleet:\\Adaptation for Large-Scale, Workflow-Based Experiments
\thanks{}
}

\newif\ifdoubleblind

\ifdoubleblind
\else
\author{\IEEEauthorblockN{Jim Pruyne}
  \IEEEauthorblockA{University of Chicago\\
    Chicago, IL\\
    pruyne@uchicago.edu}
\and
\IEEEauthorblockN{Valerie Hayot-Sasson}
\IEEEauthorblockA{University of Chicago\\
  Chicago, IL\\
  vhayot@uchicago.edu}
\and
\IEEEauthorblockN{Weijian Zheng}
\IEEEauthorblockA{Argonne National Laboratory\\
  Lemont, IL\\
  wzheng@anl.gov}
\and
\IEEEauthorblockN{Ryan Chard}
\IEEEauthorblockA{Argonne National Laboratory\\
  Lemont, IL\\
  rchard@anl.gov}
\and
\IEEEauthorblockN{Justin M. Wozniak}
\IEEEauthorblockA{Argonne National Laboratory\\
  Lemont, IL\\
  woz@anl.gov}
\and
\IEEEauthorblockN{Tekin Bicer}
\IEEEauthorblockA{Argonne National Laboratory\\
  Lemont, IL\\
  tbicer@anl.gov}
\and
\IEEEauthorblockN{Kyle Chard}
\IEEEauthorblockA{University of Chicago\\
  Chicago, IL\\
  chard@uchicago.edu}
\and
\IEEEauthorblockN{Ian~T.~Foster}
\IEEEauthorblockA{University of Chicago\\
  Chicago, IL\\
  foster@uchicago.edu}
}
\fi

\maketitle

\iffinal
\else
  \thispagestyle{plain}
  \pagestyle{plain}
\fi

\begin{abstract}
Experimental science is increasingly driven by instruments that produce vast volumes of data and thus a need to manage, compute, describe, and index this data. High performance and distributed computing provide the means of addressing the computing needs; however, in practice, the variety of actions required and the distributed set of resources involved, requires sophisticated ``flows'' defining the steps to be performed on data. As each scan or measurement is performed by an instrument, a new instance of the flow is initiated resulting in a collection of concurrently running flows, a ``fleet'', with the overall goal to process all the data collected during a potentially long-running experiment. During the course of the experiment, each flow may need to adapt its execution due to changes in the environment, such as computational or storage resource availability, or based on the progress of the fleet as a whole such as completion or discovery of an intermediate result leading to a change in subsequent flow's behavior. We introduce a cloud-based decision engine, \emph{Braid}, which flows consult during execution to query their run-time environment and coordinate with other flows within their fleet. Braid accepts streams of measurements taken from the run-time environment or from within flow runs which can then be statistically aggregated and compared to other streams to determine a strategy to guide flow execution. For example, queue lengths in execution environments can be used to direct a flow to run computations in one environment or another, or experiment progress as measured by individual flows can be aggregated to determine the progress and subsequent direction of the flows within a fleet. We describe Braid, 
its interface, implementation and performance characteristics. We further show through examples and experience modifying an existing scientific flow how Braid is used to make adaptable flows.
\end{abstract}

\section*{Keywords}
Research process automation; Globus; High-performance computing; Distributed Computing; Scientific Computing; Cloud computing

\section{Introduction}
\label{sec:introduction}


The scope of experimental science continues to increase rapidly particularly in the rate and granularity at which measurements are taken with a corresponding increase in generated data and the need to analyze, store, and archive these data. When data processing can occur concurrently with data generation, valuable insights can be 
obtained during an experiment which can, in turn, allow researchers to tune or direct experiments.
Processing experimental data as they are generated typically involves a multi-step process including data movement, computation, \Ac{ML} model training and inference, data indexing and visualization, and human-based validation. To run automatically and reliably, these steps are encoded in a ``flow'' which precisely defines these steps and which can be reliably executed and monitored by a workflow management system. 


Commonly, a flow is started for each measurement, sample or image generated by a scientific instrument. This simple, one-to-one correspondence between flow executions and experimental events may be necessary as it reduces complexity at scientific instruments where computing capacity may be low, event rates may be high, and the stakes, such as lost data, for making erroneous changes in experimental result collection processes may be high. This pattern results in many flow executions taking place during a single experiment, often running concurrently as their run-time is greater than the interval between the events which generate them. We think of this collection of flows as a ``fleet'' reflecting their independent progress with a collective goal to reach a single destination or outcome -- the result of the experiment.

When viewed as a collective fleet, we see limitations to the flow execution per experimental event approach. In particular, when running independently, it is difficult for results from early flow executions to be used by other members of the fleet to steer their progress. It is also difficult for the fleet as a whole to adapt to changes in its execution environment such as availability of compute, storage, or other resources due to the rate and automated manner in which flow executions comprising the fleet are launched. We address these inefficiencies by introducing a new service, \emph{Braid}, which can be used by flows within a fleet, as well as outside processes which may observe fleet progress or environmental conditions, to enact adaptations which individually steer flows within the fleet toward their collective goal.

Braid securely accepts and stores measurements from environmental monitors or intermediate results from flows within a fleet to provide a basis upon which adaptation decisions can be performed by running flows. The measurements may be statistically aggregated over time intervals or measurement quantity to determine trends or to eliminate outliers within the measurements. Flows may then compare multiple such aggregations to determine their strategy. For example, average waiting time for computation resources over a suitable time window may be used to determine which of many computing sites a flow would choose to run in. We present, from experience, multiple classes of flows which can take advantage of the ability to adapt to improve scientific results, the Braid service interface and examples of building flows using Braid and results from running a fleet of flows which use Braid to perform a multi-stage scientific computation.

\section{Motivation}
\label{sec:motivation}

In prior work
\ifdoubleblind
\else
~\cite{vescovi2022linking}
\fi, we described a set of production flows and their roles in various scientific experiments. These flows were used to act on data as they are collected during an experiment. 
Once started, each flow executes independently without consideration to performance or outcomes of other flows or the experiment as a whole.
Enabling communication between flows allows for interactions between the various processes and components crucial to research outcomes. 
Effective communication can ensure that data and results can be easily shared, interpreted, and validated, reducing redundancy and errors, enhancing the reproducibility and reliability of results. 
It also enables the automation of tasks, streamlining the progression from one stage of an experiment to another, which is critical for handling complex, large-scale analyses and high-throughput data inherent in contemporary scientific research. This optimization through communication is pivotal not only for accelerating scientific discoveries but also for operational efficiency, by removing redundant execution, improving utilization of instruments, and reducing time to science.



Automation significantly enhances the efficacy and reliability of both large-scale instruments, such as synchrotron x-ray sources, and smaller laboratory instruments, such as \Ac{SEM}s, providing advantages across various facets of scientific research. 
Large experimental facilities depend on automated data management practices to improve throughput, precision, and increase resource utilization by optimizing the analysis and cataloging of datasets while facilitating precise control over experimental conditions. 
In the context of smaller lab instruments, such as \Ac{SEM}s, data automation brings time efficiency, cost-effectiveness, and consistency in results. Irrespective of the scale, automation's ability to streamline experimental operations, mitigate errors, and manage high-throughput data is indispensable for the advancement of contemporary science.

Globus Flows~\cite{chard2023automate} has been applied to a variety of use cases and is routinely used to perform and automate data management and analysis tasks for scientific instruments. These flows can vary in complexity and perform a multitude of tasks, however, a common pattern involves invoking a flow on each dataset as it is generated; the flow then orchestrates the transfer of data and performs analysis on a remote resource before publishing results and indexing relevant metadata in a catalog.
A number of challenges can arise when performing such a flow, especially when data are rapidly produced and the flows leverage shared networks and compute resources. Several of these challenges are described below.


\subsection{Resource Constraints}
Facilities rarely coordinate
their availability (or unavailability) windows. Thus, for example,
an instrument facility and a computation facility may have differing maintenance schedules. In turn, the instrument may continue to run, producing data and flows to be processed, with the usual compute facility unavailable. Thus, the flows may at best be queued: eliminating the benefit of the anticipated, rapid response time of the automated flows. At worst, the
flows will fail and potential results may be lost.

Resources often have limits that restrict the number of concurrent computation jobs or data transfers. 
A basic approach to experiment automation may submit flows whenever new data is created
independent of these limits. Like in the resource unavailability situation, this may result in queuing or failures. Failures are particularly troublesome in this case because these limits are often unexpected or rarely encountered, and thus not robustly handled by flows.
  
\subsection{Experiment, Simulation, and Training Convergence}
Experiments (and simulations and model training) are often run repeatedly until results converge. Further, in some cases experiments may fail, e.g., a sample may be destroyed or machine learning may not converge. Such cases require the ability to terminate flows either when success metrics are achieved earlier than expected, or when the incremental increases are low, or when results are not meeting required goals. 

Such cases are normally represented by fleets with many flows running concurrently and a need then to send signals between flows to change their behavior. In such cases, we must determine, through policy, that an experiment is or is not converging or that individual flow runs within the fleet are erroneous. We want policy to be able to: cut short fleets that converge quickly so as not to do work that is not needed, steer back on course or abort a fleet that is headed the wrong way or re-route perhaps by re-processing to work around elements that are considered erroneous.

\subsection{Dependencies Between Flows}

As automated experiments grow more ubiquitous, they also grow more complex. We have observed processes which come in ``waves'' in the sense that data generated as output or results by one fleet becomes input and the trigger for a subsequent fleet which further processes these initial results. Thus, there is a need for methods to determine when appropriate or sufficient data is generated by the initial fleet to indicate that the subsequent fleet should be started. Commonly, this is the case when a desired state of the initial fleet is reached such that the next phase may start.

\subsection{Summary}

\begin{itemize}
\item Flows must be able to observe the state of both resources and
  the fleet they are a part of to adapt to changes in either.
\item Various conditions, especially errors that occur only under
  load, can be difficult to plan for and thus are even more difficult
  to encode in flows.
\item Adaptation modes can include routing to alternative resources,
  throttling or otherwise limiting the rate or degree to which
  resources are being used, or can involve changing the steps, such as
  the computation mode.
\end{itemize}

\section{Braid Design}
\label{sec:design}

Braid provides abstractions, delivered via a service-based REST API, to efficiently steer flow runs singularly and collectively toward a common goal. Because the scale in terms of number of flows and time over which they run can be large, and because the environment in which they run can also fluctuate and have transient states, we require that our adaptations not be overly reliant on short-term measurements or predictions. Thus, the general approach is to accumulate measurements over numerous iterations or time and make decisions based on statistical measures or trends while a fleet is performing its work. The goal is to provide a minimal set of abstractions enabling flow runs to coordinate while also making use of Braid within flows as easy as possible avoiding complex syntax or the need for sophisticated logic to be introduced into what are often otherwise simple sequences of steps.
We intend Braid's functionality to be easy to use in new flows or to incorporate into pre-existing flows at precisely the steps which would require an adaptation may guide the flow's execution.

\subsection{Concepts}
\label{sec:concepts}

We define a series of concepts which build closely upon one another to provide the functionality necessary to coordinate fleets of concurrently running flows and allow them to adapt to and interact with the environment and resources on which they run.

\subsubsection{Datastreams}
\label{sec:datastream-concept}


Any decision making capability relies on information or data. Therefore, the foundation of Braid is the container for information termed the \emph{Datastream}. Commonly, a datastream will be established to monitor a particular resource, such as expected wait time for computation, available space at (or bandwidth to) a storage site, or to monitor the progress of an experimental process as described in \autoref{sec:motivation}. A datastream is composed of a sequence of \emph{Samples} where each sample is a numeric value represents a measurement or snapshot of the state of the resource being monitored. Braid associates a timestamp with each sample, providing order which is used when performing aggregation on a datastream as described in \autoref{sec:metric-concept}.


\subsubsection{Metrics}
\label{sec:metric-concept}

While datastreams represent the information stored within Braid, \emph{Metrics} are the method for generating a single, summarization value over the samples within a datastream. Metrics are defined by
\begin{inparaenum}[1)]
\item the datastream the metric is to be computed over;
\item the \emph{operation} to be used to compute the metric value; 
\item the \emph{interval} within the datastream to compute the metric on; and
\item a \emph{parameter} value which is used by some operations to refine their result.
\end{inparaenum}

Braid supports the following set of operations for computing the metric value:
\begin{inparaenum}[1)]
  \item Average;
  \item Standard Deviation;
  \item Count;
  \item Sum;
  \item Minimum;
  \item Maximum;
  \item Mode;
  \item Continuous Percentile;
  \item Discrete Percentile;
  \item Last;
  \item First; and
  \item Constant.
\end{inparaenum}
The ``Constant'' operation always returns its parameter value which is useful in further uses of Braid, as described below, in which metric values are compared to one another. The other operations compute a value over the specified interval which may be given as a time interval or in terms of the number of samples, relative to the first and last samples in the datastream.

\subsubsection{Policy}
\label{sec:policy-concept}
  
Ultimately, the purpose of Braid is to provide decision making support for flows. A \emph{Policy} is the decision making abstraction. It consists of multiple metric specifications with the policy selecting the maximum or minimum of the computed metrics as the policy decision. Thus, metrics define the criteria for a decision, and the policy determines the preferred value among the metrics. The constant metric operation described above effectively allows a policy to compare a metric to a set value such as a threshold.

The metrics within a policy provide the criteria upon which a policy decision can be made, but they do not clearly describe how the policy decision can be used or enacted. For example, a policy may compare metrics representing the queue length at multiple compute sites averaged over the last ten minutes, but the metric does not provide information on how to access the preferred compute site. Thus, each metric in a policy request also has a \emph{Decision} value associated with it, and it is this value that Braid returns as the outcome of the policy request and which can then be used by a flow in subsequent steps. Thus, the policy's decision value can configure subsequent steps of a flow directly, without branching or other logic which often adds significant complexity to the flow development process. Furthering the example of selecting among compute sites, the decision values associated with the metrics would include identifiers, addresses, or other configuration information specific to using the selected compute site. 

Often, a datastream is configured to monitor a particular resource so properties of that resource can be configured as a ``default decision'' on a datastream so that a policy specification can omit the decision value on any metrics involving datastreams which define a default decision. In this way, the creator of a datastream who likely has most direct knowledge of the resource being monitored or measured by the datastream can provide this information within the datastream definition and then future users, such as those writing policy statements into their flows, need not be familiar with these specific details and use the values provided by the datastream creator.

\subsection{Implementation}
\label{sec:implementation}

Braid is deployed as a scaleable, cloud-hosted service making it suitable for use by production scientific experiments. The datastream, metric and policy concepts described above are accessed through a REST-style programming interface. Datastreams are the foundation resource in the REST model. Each datastream is given a descriptive name at creation and the service generates a unique identifier which is used to reference the datastream in all subsequent operations including the datastream's life-cycle, adding samples and evaluating metrics and policies.

\subsubsection{Authorization}
\label{sec:authorization}

As an internet accessible service, Braid must also provide protection for data it stores. As a datastream is referenced in all operations including adding samples and evaluating metrics and policies, the datastream also provides the primary abstraction through which authorization is performed via a set of roles associated with it. The user who creates the datastream is assigned the \emph{Owner} role, and they may update the datastream to change the name, and the identity of users in other roles (including the owner role allowing an owner to transfer ownership to another user). The \emph{Provider} role governs which users are allowed to add samples to the datastream. The \emph{Querier} role likewise defines which users are allowed to access the datastream to evaluate metrics and policies. These roles provide a valuable separation of concerns. For example, a user may establish a datastream to monitor a resource, grant a provider role to a user with access to monitor the resource and make the datastream information available to a collection of other users, those running flows which need to access the datastream, by granting the querier role to them. In this scenario, neither the owner user nor any of the querier users need be concerned with malicious or erroneous samples entering the datastream as long as the provider is reliable and trusted.

All authentication operations are performed via Globus Auth~\cite{GlobusAuth} which provides an OAuth-2~\cite{oauth2} interface for the generation and validation of tokens associated with each REST API request. Further, identities associated with Globus Auth users may also be placed in Globus Groups~\cite{chard16nexus} and thus roles can be assigned to these groups allowing a changeable set of users to be associated with any role without making updates directly updating Braid to enumerate the members.

\subsubsection{Client Tools}
\label{sec:client-tools}

We provide client tooling in the form of a \Ac{CLI} and a Python \Ac{SDK} which can each be used to access the full set of service interfaces. The \Ac{CLI} is typically used when setting up an experiment environment such as creating and setting roles on datastreams or providing initial samples to set state needed when flows begin running. The \Ac{SDK} provides a means of accessing the Braid service in more complicated or dynamic ways often involving interacting with resources and providing on-going sampling of their states. We show examples of the use of the \Ac{CLI} and \Ac{SDK} in \autoref{sec:usage}.

\subsubsection{Flow-specific interfaces}
\label{sec:flow-interfaces}

In addition to the typical REST interface as supported by the above tools, Braid's principal use case for use within flows requires a method for it to be invoked from a flow. We run our flows on the Globus Flows service, and therefore Braid implements the ``Action Provider'' interface~\cite{chard2023automate}. 
In particular, the flows are able to invoke the following Braid functions: 
\begin{inparaenum}[1)]
\item adding samples to a datastream;
\item evaluating a policy; and
\item waiting for a policy to reach a particular decision.
\end{inparaenum}
Each of these operations within a flow is subject to the same authorization requirements: to add a sample, the user running the flow must be a provider to the datastream and to evaluate a policy, the flow running user must be a querier of the datastream.

The ability to not just evaluate, but also wait for a policy decision (the last operation in the previous list) provides a commonly used method for synchronizing flows without need for loops, retries, back-offs or timeouts to be encoded in a flow's logic which is highly error prone given the care needed to encode these operations correctly and the limited expressive capability of flow syntax. The use of these operations within flows will be shown in more detail in the example in \autoref{sec:usage}.

\section{Usage}
\label{sec:usage}

Here, we present a detailed example of a fleet of flows adapting their execution and coordinating their execution using Braid. The example presented is similar to results presented in \autoref{sec:experience} though somewhat simplified to illustrate the use of Braid. The details presented here will also be applicable when understanding the results presented later. The scenario illustrated here can be summarized as follows:
\kyle{Would be good to make this a bit more specific, maybe even use the example from later? or part of a real example?}
\begin{quote}
A computational experiment consisting of a number of flows, containing independent computations, are set to run concurrently. Each computation creates a result as well as an estimate for the progress of the experiment as a whole. When a sufficient number of progress estimates indicate that the computation is completed, each flow must also execute a ``finalization'' computation correlating its initial result with the overall result computed across the flows.

Each flow may run using either of two compute clusters and the availability of capacity within the clusters may fluctuate during the lifetime of the experiment so the environment to be used should be selected at the start of each flow rather than at the start of the experiment as a whole.
\end{quote}

In the scenario, we use three Datastreams for evaluating these two Policies:
\begin{inparaenum}[1)]
\item When has the experiment made sufficient progress that finalization can be computed by comparing a single datastream to a constant, threshold value; and
\item For each computation, which of the two clusters should be used by comparing recent values within datastreams monitoring each of the clusters.
\end{inparaenum}
To run the complete experiment, Braid must be used in the following manners:
\begin{inparaenum}[1)]
\item in an ``administration'' capacity, via the \Ac{CLI}, to setup initial state to be used during the experiment;
\item in a ``monitoring'' mode, via the \Ac{SDK}, in which samples representing the capacity of the clusters are periodically sent to the appropriate datastreams; and
\item within flow execution when Braid will both be consulted for policy decisions and updated with samples representing the progress of the experiment.
\end{inparaenum}
Each of these interaction methods is outlined below.

\subsection{Administrative Usage}
\label{sec:usage:admin}
\Ac{CLI} usage to create a datastream takes the following form:
\begin{lstlisting}[language=bash, caption=Datastream creation via the CLI Interface,label=listing:datastream-creation]
policy datastream create
  --name cluster1
  --providers <monitoring_user_id>
  --queriers <experiment_runner_user_id>
  --default-decision '{"cluster_id": "<id_val>"}'
\end{lstlisting}

The parameters to this command define all of the properties for the datastream to be created including:
\begin{inparaenum}[1)]
\item the human readable name that will be presented when datastream information is presented later (for example via a \texttt{datastream list} sub-command);
\item the ``providers'' user(s) who can provide samples into the datastream (e.g. via the programmatic usage described next)
\item the ``queriers'' user(s) who can retrieve information from the datastream including via policy evaluations (e.g. within a flow execution demonstrated below); and
\item the ``default decision'' value returned from a policy evaluation when a metric referencing this datastream is selected in a policy evaluation.
\end{inparaenum}
Specifying the default decision here allows the administrator, who is most likely to know the details needed to access this cluster, to provide that information so various flows which may use the cluster need not each have these details embedded.

\subsection{Programmatic Usage}
\label{sec:usage:monitoring}
Monitoring the cluster could be performed with a daemon process written using the \Ac{SDK} such as the following:

\begin{lstlisting}[language=bash, caption=Populating a Datastream via the SDK Interface]
client = BraidClient()
while True:
  availability = get_cluster_availability()
  client.add_sample(datastream_id,
                    value=availability)
  time.sleep(sample_interval)
\end{lstlisting}

This code snippet starts by creating an instance of the \Ac{SDK}'s client class, and then enters its monitoring loop. Each iteration of the loop gets an estimate for the ``availability'' of the cluster and then sends it as a sample to the configured datastream. Finally, the monitor sleeps for some interval such that updates are sent at a regular interval. In our example scenario, we have two available clusters, so two instances of this monitoring script may be run. These scripts should continue to run throughout the experiment so that updated availability information is always available to the flows.

\subsection{Flow Execution Usage}
\label{sec:usage:flow-execution}
The administrative and programmatic interactions set up use by each of the many flows within the fleet that makes up the experiment. The flow implementing our example scenario comprises five steps, three of which are invocations of Braid. The step are described below in the sequence they appear in the complete flow. We use a syntax based on the Amazon States Language~\cite{amazonstateslanguage} with the inclusion of a new property ``ActionUrl'' which makes the step perform a call out to the service at the corresponding URL which is either the Braid service or a canonical compute service used in the example.

\begin{lstlisting}[language=json]
{
  "ActionUrl": "<braid_location>/policy_eval",
  "Parameters": {
    "metrics": [{
      "datastream_id": "cluster_monitor_id_1",
      "op": "avg"}, {
      "datastream_id": "cluster_monitor_id_2",
      "op": "avg"
    }],
    "policy_start_time": -600,
    "target": "max"
    }
  "ResultPath": "$.PolicyDecision"
}
\end{lstlisting}

The first step of the flow uses Braid to evaluate a policy determining which cluster will be used throughout the run of this flow instance. The metrics to compare are from the datastreams populated programmatically above (\texttt{cluster\_monitor\_id\_1}, \texttt{cluster\_monitor\_id\_2}) computing the average (\texttt{avg}) for each over the last ten minutes (\texttt{policy\_start\_time} value \texttt{-600} seconds) to select the maximum value. The \texttt{ResultPath} specification causes the policy decision output to be stored in the state of the flow under the key \texttt{PolicyDecision} and, as configured in the administrative section, the decision will include a value for \texttt{cluster\_id} indicating which cluster, not just datastream, is considered to have the best availability to run the flow's computations most rapidly.

\begin{lstlisting}[language=json]
{
  "ActionUrl": "<compute_service location>",
  "Parameters": {
    "cluster_id.$": "$.PolicyDecision.decision.cluster_id",   
    "comptation_parameters": {...}
  },
  "ResultPath": "$.ComputationResult"
}
\end{lstlisting}

The second step of the flow runs the primary computation associated with the flow via a computation service. The computation service invocation is parameterized with the \texttt{cluster\_id} of the location on which the computation should run. This is set referencing the value returned in the decision of the previous, policy evaluation step using a JSONPath reference (as defined by to the Amazon States Language). Per the scenario definition, the computation result will contain a ``quality'' value which will be stored under the \texttt{ComputationResult} key in the flow's state.

\begin{lstlisting}[language=json]
{
  "ActionUrl": "<braid_location>/add_sample",
  "Parameters": {
    "datastream_id": "result_quality_datastream_id",
    "value.$": "$.ComputationResult.result_quality"
  },
  "ResultPath": "$.SampleResult"
}
\end{lstlisting}

The quality of the computation is next populated into the datastream created for monitoring the progress of the fleet as a whole. This is done using the \texttt{add\_sample} flow interface. The value populated is retrieved from the flow's state under 
the \texttt{\$.ComputationResult.result\_quality} key created in the previous step.

\begin{lstlisting}[language=json]
{
  "ActionUrl": "<braid_location>/policy_wait",
  "Parameters": {
    "metrics": [{
        "datastream_id": "result_quality_datastream_id",
        "op": "discrete_percentile",
        "op_param": 0.9,
        "decision": "wait"
      },{
        "datastream_id": "result_quality_datastream_id",
        "op": "constant",
        "op_param": 0.95,
        "decision": "proceed"
      }],
    "policy_start_limit": -10,
    "target": "min",
    "wait_for_decision": "proceed"
  },
  "ResultPath": "$.WaitPolicyDecision"
}

\end{lstlisting}

The previous step sampled the result quality to Braid and must wait for other flows within the fleet to also report quality results and, further, for those collected results to meet a threshold value before each flow may proceed. The policy wait flow interface, introduced in \autoref{sec:flow-interfaces}, serves this purpose by holding a flow blocked until the policy evaluation matches a desired decision value (the underlying flow syntax allows for setting maximum step run times and handling timeout exceptions should a policy wait hold the flow for an unexpectedly long time). The condition to be met is ``9 out of the last 10 quality samples must have a value of at least 0.95'' which is expressed by:
\begin{inparaenum}[1)]
\item evaluating a metric on the quality datastream to compute the 90th (\texttt{0.9}) percentile of the samples and setting the decision value to ``wait'' if this metric meets the policy criteria;
\item evaluating a second metric using a \texttt{constant} operation set to the value of 0.95 with a decision value of ``proceed'';
\item specifying that the metrics should be computed over the ten most recent samples (\texttt{policy\_start\_limit} value -10);
\item that the target criteria should be the minimum of the computed metrics and thus if the constant is less than the computed quality its decision value ``proceed'' is selected; and
\item the flow will remain paused at this step until the decision has the value ``proceed.''
\end{inparaenum}

\begin{lstlisting}[language=json]
{
  "ActionUrl": "<compute_service location>",
  "Parameters": {
    "cluster_id": "$.PolicyDecision.decision.cluster_id",
    "comptation_parameters": {...}
  },
  "ResultPath": "$.FinalizationComputationResult"
}
\end{lstlisting}

The last step of the flow is another computation using the same cluster as the earlier computation. At this step of the flow, both this flow's quality and the aggregate quality computed in the policy of the previous step are available parameters to the computation. Thus, it could calculate a final result comparing its result to the aggregate or it could simply store its relative results in a filesystem.

The flow presented here encapsulates some of the motivating concepts described in \autoref{sec:motivation} and is similar to the practical, science based application described in \autoref{sec:experience}. While concepts such as resource selection, experiment progress measurement and synchronization of flows within a fleet are each somewhat complex, the flow itself is fairly simple to read and to code. It involves no conditionals and no loops: constructs which are more prone to errors especially when implemented by non-programmers who are often the target audience for workflow systems often promoted as ``no code'' solutions.

\section{Evaluation}
\label{sec:evaluation}
Braid is deployed as a cloud-hosted service on \Ac{AWS} using a variety of capabilities for deployment, protected interfacing to the internet, flexible execution environments, and flexible database back-ends. In particular, we employ the \Ac{AWS} \Ac{ECS}~\cite{AmazonECS} to host the service and Aurora~\cite{AmazonAurora} running Postgres~\cite{Postgres} for the database layer. The service is implemented in Python using the FastAPI~\cite{FastAPI} framework and packaged as a Docker container for deployment to \Ac{ECS}. Use of flexible cloud hosting means that we can easily scale up or down the capacity of the service based on load and cost constraints.

To obtain a baseline for the performance and scaleability of the service, we established a deployment in \Ac{ECS} with two service instances, each with 8GB of memory and 2 \Ac{vCPU}s. The Aurora database cluster is provisioned to scale to a maximum of 4 \Ac{ACU}s. In production use, we impose rate limits on samples ingested as well as metric and policy evaluations performed. To constrain storage consumption, we
cap the total number of samples retained in any one datastream to one million entries with older entries automatically removed.

\subsection{Performance Testing}

We employ a series of micro-benchmarks to evaluate both the rate at which samples can be ingested and the rate at which metric evaluations of various types can be completed. In each of the tests, the clients which perform the measurement are hosted in \Ac{AWS} to minimize the variability introduced by internet wide latency. 

\begin{figure}
    \setlength{\fboxsep}{0pt}
    \includegraphics[width=0.5\textwidth]{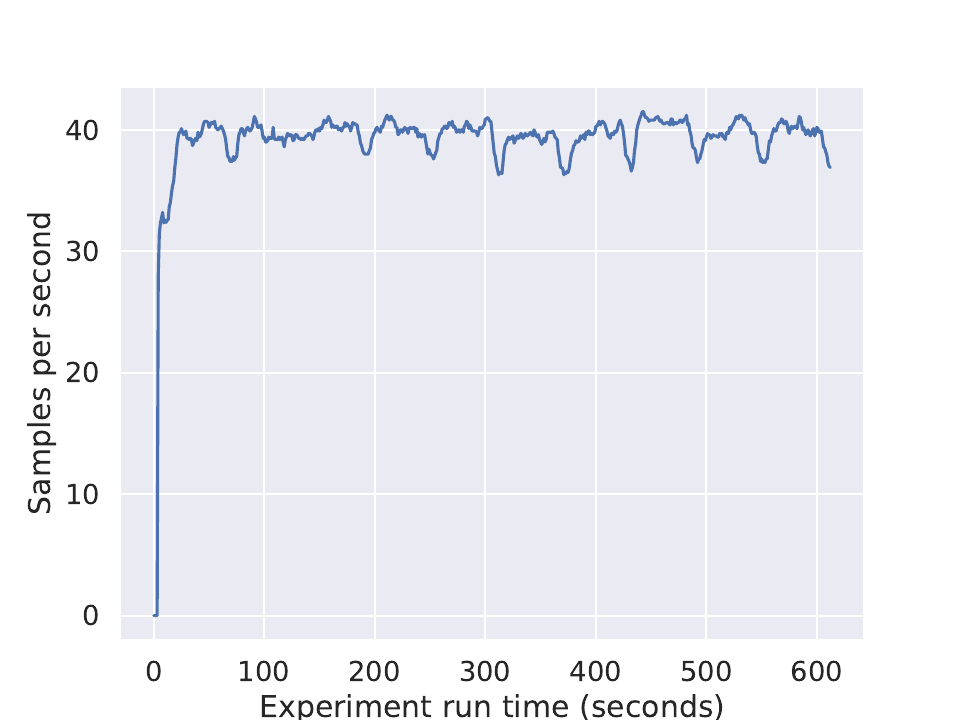}
    \caption{Request rates driven by a single client}
    \label{fig:datastream}
\end{figure}

The first micro-benchmarks measures the rate at which samples can be ingested into a single datastream by a single client which waits for completion of each request before initiating the next. We implement this benchmark using Locust~\cite{locust}, a Python performance testing tool. \autoref{fig:datastream} shows that in this scenario, Braid can process an average of 37 and a maximum of 41 requests per second with no significant degradation as the datastsream grows. We do see, periodic, sharp dips in request rate. These  can be attributed to Braid periodically re-validating the credentials presented by the client making the request. This validation requires a remote call to the authorization service so adds signficant time to what the client observes as the rate for that request.


\begin{figure}
    \centering
    \includegraphics[width=0.5\textwidth]{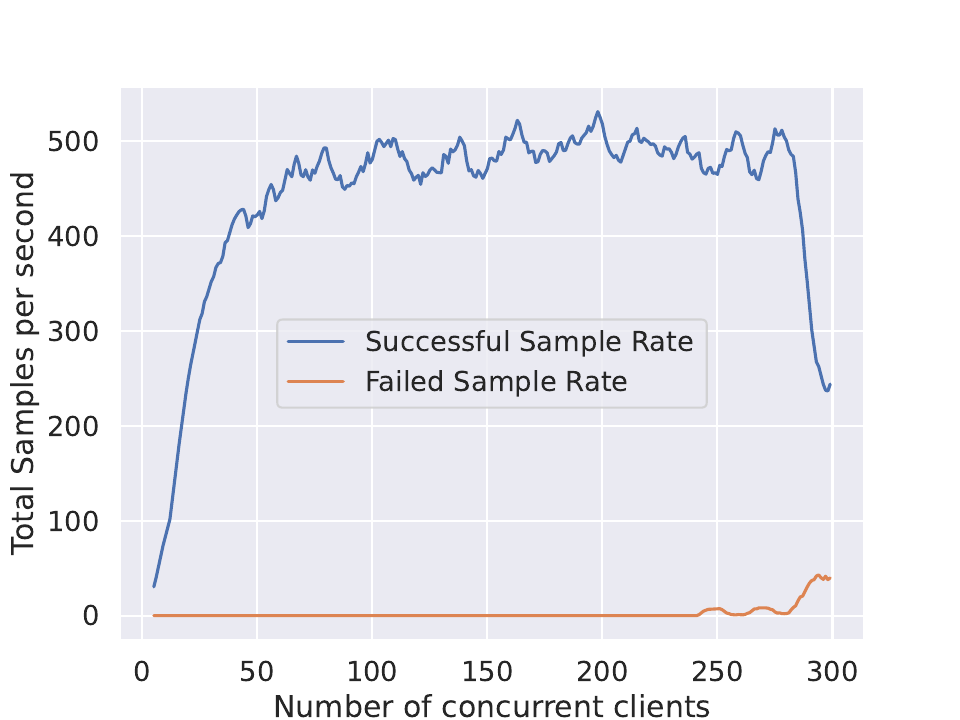}
    \caption{Request and failure rates driven by multiple, concurrent clients}
    \label{fig:manyworkers}
\end{figure}

Our second micro-benchmark measures how the service behaves under concurrent load from multiple clients each adding samples to a separate datastream. Once again, we use Locust to perform this evaluation. Our results (Figure~\ref{fig:manyworkers}) demonstrate that the service can deliver a peak of just over 500 requests per second with a sustained mean rate of approximate 470 requests per second. We again see the saw-tooth shape due to periodic requests for authenticating validation credentials. As we approach 250 concurrent clients, we begin to see errors due to clients timing out waiting for a result and at about 270 clients the error rate increases dramatically as server-side connection pools begin to fill up causing client connections to be dropped. We view these results as inputs to help us set reasonable rate caps for our clients and guidance on setting the size of the service we provision in \Ac{AWS}. We see that the moderately sized service cluster we provisioned for this experiment would be able to handle signficant load.

\begin{figure}
    \setlength{\fboxsep}{0pt}
    \includegraphics[width=0.5\textwidth]{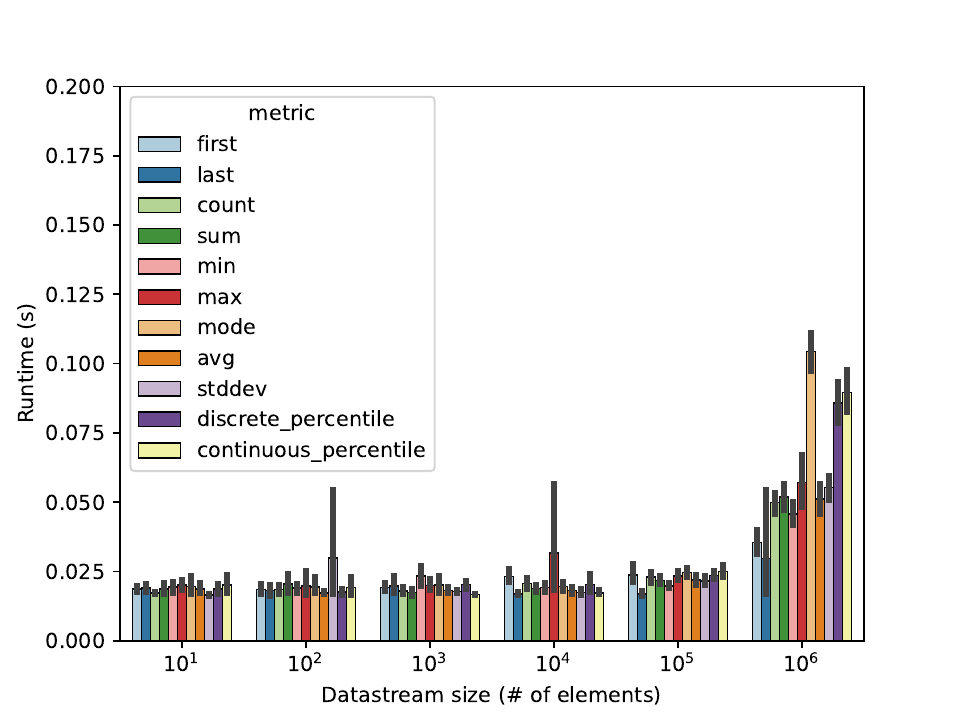}
    \caption{Metric performance across various datastream sizes}
    \label{fig:metrics}
\end{figure}

Finally, we want to know the time required to evaluate the various metric operations (as enumerated in \autoref{sec:metric-concept}) for various size datastreams. We need to verify that subject to the limits we impose, metric computations can be completed in the time needed to perform adaptation for our flows. We initialized Braid with datastreams of sizes ranging from 10 to 1,000,000 entries. We then selected a datastream and a metric operation combination to be computed and repeated until each combination had been computed at least 10 times. We perform this random combination to get worst-case results, avoiding caching or other optimizations in the database that might be caused by repeating the same operation consecutively. \autoref{fig:metrics} shows the result of these tests. The most important result is that, even for datastreams of size 1,000,000, any metric can be computed in no more than about 100 milliseconds. Outliers do show up and we attribute these to network delay as a small change in network latency has a signficant impact on transaction times measured in the 10's of milliseconds.

We attribute the generally good performance to Braid's implementation the metrics: each is computed using a single SQL query using a built-in function or, in the case of \texttt{First} and \texttt{Last} a combination of SQL ``ORDER BY'' and ``LIMIT'' clauses. Operations common to flows such as data transfer and computation typically take a minimum of a few seconds, and often minutes or longer, so these times to compute metrics would not have a significant effect on overall flow execution times.

\section{Experience with High Energy Diffusion Microscopy}
\label{sec:experience}

High-energy X-ray diffraction methods allow scientists to non-destructively analyze bulk metallic polycrystalline engineering materials. Specifically, the \Ac{HEDM}~\cite{pokharel18hedm} technique can extract three dimensional microstructure information and grain-specific properties. \Ac{HEDM} produces digital scans of a material's evolving microstructure and its associated attributes, enabling scientists to observe and understand how materials transform when external stimuli, such as increased pressure load, are applied~\cite{naragani.2017rp}. 

Automated computational methods have been developed that can provide near-real-time feedback to scientists.
In particular, unsupervised learning has been used to identify anomalies within X-ray scans as they are collected. The technique works as follows: prior to the experiment starting, scientists identify a key condition, such as the pressure on the material, which represents a \emph{baseline} upon which training will be performed to generate cluster centers. Subsequent to the training, each scan, including those collected prior to the baseline condition, are assessed relative to the cluster centers to calculate an \emph{anomaly score}. The domain scientists also determine a threshold for the anomaly scores which would imply that the microscopic indicator has occurred and thus the experiment can be concluded.

We encode the automatic anomaly detection process in two flows: a training flow that is run only once, when the baseline scan is collected, and a second which is run for every scan to compute its anomaly score. We start the anomaly score flow as each scan is collected, but the computation of the score cannot actually be performed until training on the baseline scan is complete.

We track the experiment through three phases:
\begin{inparaenum}[1)]
\item scans have been collected, but the baseline scan has not yet been collected, so anomaly score calculations cannot yet be computed and the flows for these scans must wait until the flow that performs the training has completed;
\item training has been completed, and thus waiting flows can progress, using the model created by training, and all incoming scans can have their anomaly scores computed immediately; and
\item a sufficient quantity of scans have produced the desired anomaly score, and thus the experiment can be considered complete and no subsequent scans need to be taken.
\end{inparaenum}

These two flows use two datastreams to guide the experiment through the three phases. A ``coordination'' datastream tracks the phase of the experiment. When this datastream is created during administrative experiment setup, an initial sample value of 1.0 is ingested establishing that the experiment is in phase one. The second datastream receives samples for each of the anomaly scores when they are computed and is then consulted by each flow to determine if the desired anomaly score threshold has been reached and thus the experiment will move to phase three indicating completion. The steps of the flows are as follows:

\begin{lstlisting}[language=numberedflow,caption=Steps in the ``Training'' flow]
Transfer scan data from instrument to computer
Perform Training
Add sample value 2.0 to coordination datastream
\end{lstlisting}

\begin{lstlisting}[language=numberedflow, caption=Steps in the ``Anomaly Score Computation'' flow]
Transfer scan data from instrument to computer
Wait for policy on coordination datastream to be at least 2.0
Perform anomaly score computation
Sample computed anomaly score to anomaly score datastream
Evaluate policy to determine updated experiment phase based on aggregate Anomaly score values
Sample policy output to the coordination Datastream
\end{lstlisting}

In our environment, data (scan) collection and computation take place at different sites, requiring the first step of each flow to transfer data from the instrument to the computation site. In the training flow, the subsequent steps are to perform the training computation, which takes on the order of minutes, and then publish the constant value 2.0 to the Coordination datastream indicating the move to the second phase. The model generated by the training is written to the filesystem at the computation site so that it can be used by the anomaly score computation.

The anomaly score computation is similar to the example shown in \autoref{sec:usage}. It starts with a data transfer step, and then immediately uses a policy wait step to hold until the coordination datastream has a value of at least 2.0 indicating that the training is complete. This ordering allows the transfer to be completed even if training has not yet completed such that when training is complete, the flow can continue immediately. The next steps are similar to the example: computation, followed by sampling the result of the computation followed by a policy evaluation over the recent samples. Here, the result of the policy is published directly to the coordination datastream. The condition for the policy is written by the domain scientists based on their knowledge of the anomaly score calculation and the experiment being conducted. In the experiment shown here, the policy is the same as shown in the example of \autoref{sec:usage}: ``9 of the last 10 anomaly scores with a value greater the 0.95''. In this case, the decision value for meeting this policy is set to 3.0 such that when it is sampled to the coordination datastream, we indicate that the experiment as a whole has reached phase 3: completion.

\begin{figure}
    \setlength{\fboxsep}{0pt}
    \includegraphics[width=0.5\textwidth]{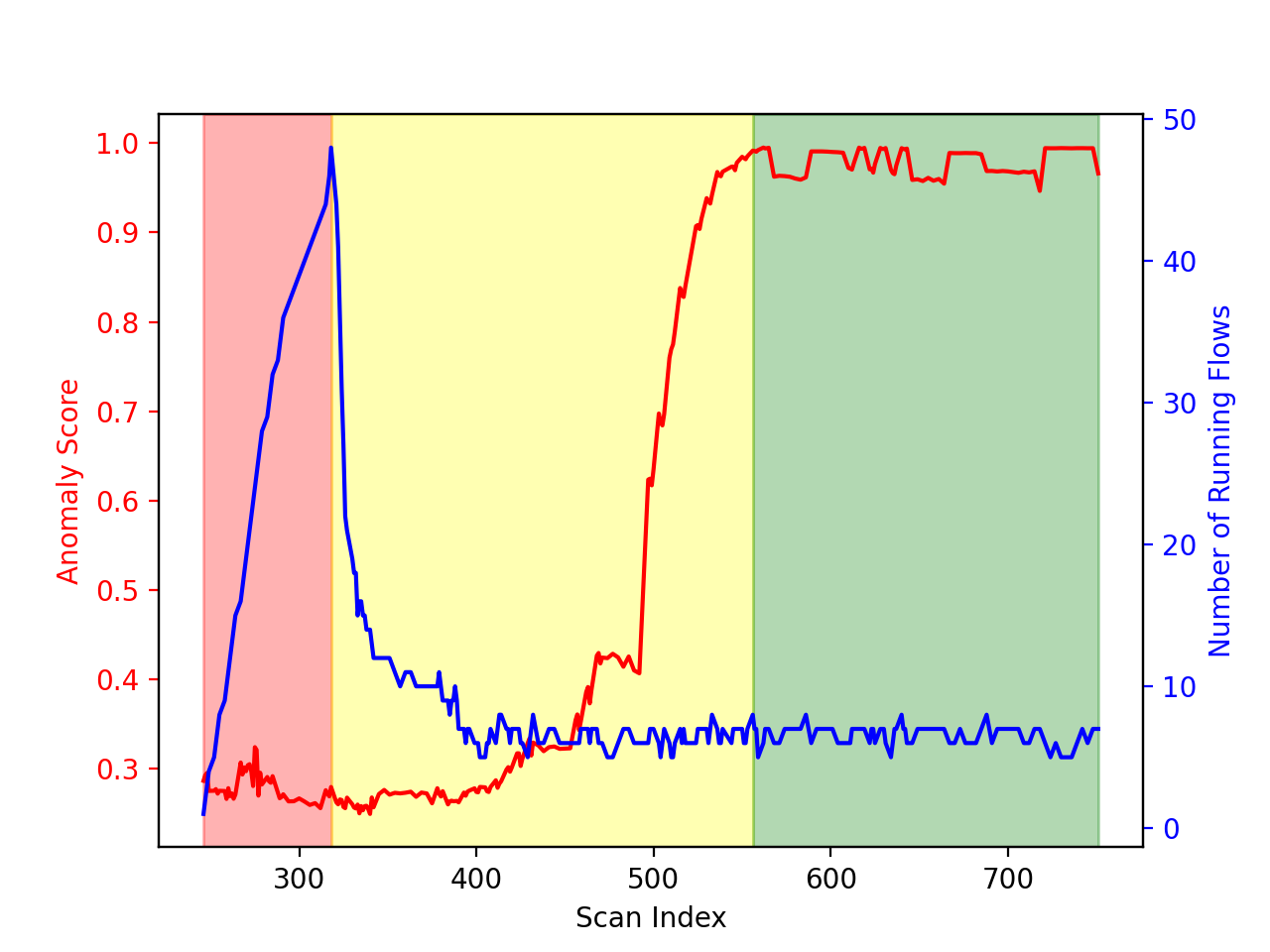}
    \caption{Progress of the HEDM experiment}
    \label{fig:hedm-progress}
\end{figure}

\autoref{fig:hedm-progress} shows the results of running the flows described above over a dataset comprising 262 individual scan files. Each scan is assigned an integer \emph{scan index}. The scan index value increases as the experiment progresses; in this experiment, from 246 to 751. However, the experimental process does not increment the index value uniformly over time. That is, there is no set interval between the assigned scan index values. Thus, while the horizontal axis of the figure is the scan index, the values plotted are not uniformly distributed across the horizontal access. We emulate a run of the experiment by iterating over each of the scan files with a delay of 10 seconds between each. For each scan file, we start an instance of the anomaly score computation flow. When the baseline scan, which is determined by the domain scientists to be index 318 in this dataset, is selected, the training flow is started.

The shading of the figure indicates the phase of the experiment. The red background is phase 1, prior to completion of the training flow. Phase 2 is indicated by the yellow background, and phase 3 is the green background. The blue line indicates the number of anomaly score computation flows running at the time the flow for that scan index is started. We see that the number of concurrent flows increases during phase 1 as no flows can complete until training has completed. Upon entering phase 2, the waiting flows complete quickly and the experiment reaches a relatively steady state in which between five and eight flows are active until the end of the experiment. The red plot presents the anomaly score calculated for each scan. At scan index 556 the experiment completion policy is achieved and transition to phase 3 is signaled by that flow sampling the 3 value to the coordination datastream. Phase 3 continues for an additional 81 scans, indices 557 through 751, which are unneeded based on the anomaly score criteria being met indicating that the desired transition of the material under test has been observed. Thus, these 81 scans, representing about 30\% of the total taken during the experiment could have been avoided. This, in turn, can free up the apparatus and the scientists to move on to further experiments yielding more results in the same amount time and resource allocation, a large potential benefit when access to large scale scientific instruments is scarce.

\section{Related Work}
\label{sec:related}

Braid provides capability similar to many other systems in the parallel and distributed processing space while our focus particularly on coordination of workflows via advanced policy also leads to differences.  Other systems with rich, high-level policy goals are commonly oriented around authorization mechanisms, but we include some illustrative examples here. Thus, the related work for Braid consists of wide area workflow control, policy specification formats, policy management databases, and policy rule engines.

\subsection{Workflow control}

Wide-area, resilient distributed computing is a challenging control problem. Adaptive approaches in a GRAM~\cite{GRAM_1998} or Condor~\cite{Condor_Experience_2004} include FTSH~\cite{FTSH_2003}, which enabled scripted replication and retry in the presence of a range of failure modes, including soft failures due to congestion.  Handling exceptions in workflows explicitly has been studied~\cite{Workflow_Exception_2006}, an approach that requires programmer attention.  Replicated workflows~\cite{Workflow_Replication_2021} have continued to be studied on commercial and open cloud infrastructure.

Our approach is based on previously developed technologies, which include distributing computation via Globus Compute (funcX)~\cite{chard20funcx}, transferring files via Globus Transfer~\cite{allen12software,liu2021design}, managing data permissions via Globus Share~\cite{chard2014efficient}, and managining persistent identifiers via Globus Identifiers~\cite{ananthakrishnan20identifiers}.  Relevant alternative approaches include distributing work via SSH~\cite{alt20oauthssh}, but the high-level logic of the workflow must be encoded somehow to allow high-level policies to adapt and coordinate the workflow execution.  Run-time adaptivity was explored in Pegasus~\cite{Pegasus_Adaptive_2009}, an approach in which workflow schedules are adapted over time as performance data on the underlying resources becomes available.

\subsection{Policy specification formats and languages}

Policies must be specified by a user, human or otherwise.  Multiple traditions exist for policy specification, including highly collaborative, user friendly specification models from the business operations space, and technical specifications often based on existing structured file formats from the database and data management space.

In Web services standards there has been an effort to create rich XML-based languages defining policies that determine access control. Powerful ontological languages such as the Web Ontology Language (OWL)~\cite{OWL_2012} have gathered support as they are more powerful in terms of semantic expressiveness and provide simple reasoning capabilities.

eXtensible Access Control Markup Language (XACML)~\cite{ferraiolo2016comparison} is an OASIS standard that defines an XML based architecture providing functionality to create, evaluate and enforce policies. In addition to the access control policy language it also defines a request/response language. XACML enables the use of a decision point to store and evaluate user defined policies. XACML policies are customizable and therefore facilitate user defined functions, making it well suited for DRIVE as policies must be highly customizable and able to be applied in different domains (provider types). The Policy Decision Point (PDP) makes the decision whether or not to authorise a request and the Policy Enforcement Point (PEP) enforces the policy decisions.


Enterprise Privacy Authorization Language (EPAL)~\cite{ashley2003enterprise} is a XML-based functional language developed by IBM to represent enterprise security policies. EPAL is very similar to XACML in most respects including representation, functionality and policy enforcement model. A comparison of features of the two found that EPAL supports a subset of almost all features included in XACML~\cite{anderson2006comparison}.

Ponder~\cite{damianou2001ponder} is a policy language designed specifically for management and security of distributed systems. Ponder policies are described using an extensible object oriented model which allows definition and multiple instantiations of policies. The language however is designed to specify general policies with little interdependence and the language itself is somewhat complex.

\subsection{Policy management databases}

Following specification, policies must be encoded into persistent storage and accessible via useful queries.

KAoS~\cite{johnson2003kaos} uses OWL to represent policy.  Designed for operation inside a Tomcat web server, accepts policy creation specifications from users using web-based tools.  KAoS encodes the policies into OWL using Jena, hiding complexity and simplifying the user experience.   KAoS then distributes the policy to the distributed JVMs that are acting as PDPs.

Relational databases have been conceptually extended to contain record-level privacy constraints, for example by IBM~\cite{DB_Privacy_2005}.  This effort considered the translating a P3P~\cite{cranor2002platform} policy from XML to an internal restriction specification.  The restrictions are capable of managing intersections and unions.  The approach was limited to managing the database itself, does not seem to have an implementation, and does not seem to satisfy Codd's rule 1 (representing all data in tables).

MirAIe~\cite{MirAIe_2023} uses a shared cache such as Redis.  This cache is responsible for recording the state of various devices triggers and caching trigger-related from sensors, avoiding recomputing the state of all triggers and devices.

\subsection{Policy rule engines}

Finally, policies must be evaluated at time of use.  Generally, rule engines are commonly used in business processing systems, which process some variation of the Rete algorithm~\cite{Rete_1982} to process large numbers of rules.

Rule based engines such as Jess~\cite{Jess_1997} are also commonly used to express policies in Java applications.  Jess was designed to support expert systems by evaluating user-defined rules over a user-defined data.  Following the Rete algorithm, rules may be triggered when their conditions are satisfied, including the execution of Java code.  Drools~\cite{browne2009jboss,drools2015} is a business rules engine that also implements the Rete algorithm and integrates with an enterprise application suite and JSR~94~\cite{JSR_94}.  Distributed Drools evaluation has been performed in Spark~\cite{Spark_2010} by placing the Drools engine inside an RDD map function~\cite{Drools_Spark_2019}.  The IBM Operational Decision Manager~\cite{IBM_Decision_Manual} offers a visual rule definition and execution programming model, and also integrates with enterprise services.

Rei~\cite{kagal2002rei} is a Prolog-based policy engine that supports Prolog and RDF~\cite{brickley1998resource} based descriptions, and evaluates them using a Java engine.  Rei contains a rich set of relationships among Subjects and their Rights, Obligations, and Delegations.

The MirAIe rule engine~\cite{MirAIe_2023} is capable of responding to sensors in an Internet of things context.  The outputs can include calls to web services or messages issued via Kafka~\cite{Kafka_2013}.  MirAIe includes a horizontally scalable trigger processing service, deployed in Kubernetes~\cite{hightower17kubernetes}, that is reliable through the use of reliable messages and a replicated underlying trigger cache.

Many existing systems contain embedded rule engines that serve the purpose of that system.  iRODS~\cite{iRODS_2006,moore2007rule} is a storage management system that enables access to distributed storage resources through a partially centralized service.  iRODS uses a rules-based paradigm for self-management, including the expected user access rules for access control, and also rules for data placement, migration, parallel access, and consistency levels.  PolicyCop~\cite{PolicyCop_2013} monitors network traffic and validates it against rules specified by the user in a database; rules violations may trigger changes in the underlying Software Defined Network (SDN) being managed by the system.

\section{Conclusions}
\label{sec:conclusion}

As instrument-based science increasingly relies on computing and data management services to process the output from these devices, reliable, scaleable computing paradigms such as flows, and fleets of concurrent flows become essential to completing these experiments. Further, as the fleets themselves grow and the flows become more complex, we have discovered that additional services are needed to steer these experiments toward their desired results. The Braid service provides the capabilities needed to add steps into these flows, building on measurements both within the fleet of flows and from their surrounding environments, such that the flows can make these dynamic adaptations. Braid has been shown to scale to the degree necessary to support these many flows and the abstractions and service interfaces have been used to create flows not previously possible which improve the efficiency of existing scientific processes. We operate Braid as an always-on, cloud hosted service continuously available for production usage. We anticipate that as the variety of experiments running through flows expands we will continue to encounter more opportunities to support these flows with additional services with the continued goal of increasing the rate and reliability with which these types of data-driven experiments can be completed.

\iffinal
\jim{Braid grant information here}
\fi

\balance

\bibliographystyle{ieeetr}
\bibliography{refs}


\end{document}